\documentclass[%
 reprint,
 amsmath,amssymb,
 aps,
 pra,
 superscriptaddress
]{revtex4-1}

\usepackage{graphicx}
\usepackage{dcolumn}
\usepackage{bm}
\usepackage{physics}
\usepackage{fullpage}
\usepackage{fancyhdr}
\usepackage{mathtools}
\usepackage{listings}
\usepackage{amssymb}
\usepackage{bbold}
\usepackage{array,multirow}
\usepackage{hyperref}

\usepackage{wrapfig}
\usepackage[caption=false]{subfig}
\newcommand\figref{Figure~\ref}

\usepackage{color}   
\usepackage{hyperref}
\hypersetup{
    colorlinks=true, %
    linktoc=all,     %
    linkcolor=blue,  
}

\usepackage{booktabs}

\global\long\def\Z{\textnormal{\textbf{Z}}}%
\global\long\def\U{\textnormal{\textbf{U}}}%

\global\long\def\ph{a^{\dagger}a}%
\global\long\def\phm{b^{\dagger}b}%
\global\long\def\bb{b}%

\global\long\def\hb{H_{\text{bos}}}%
\global\long\def\vb{V_{\text{bos}}}%
\global\long\def\dg{\dagger}%

\begin{document}


\title{Approximating the two-mode two-photon Rabi model}

\author{David H. Wu}
\email{dhwu@caltech.edu}
\affiliation{Institute for Quantum Information and Matter \& Walter Burke Institute for Theoretical Physics, California Institute of Technology, Pasadena CA, USA}
\author{Victor V. Albert}%
\affiliation{Institute for Quantum Information and Matter \& Walter Burke Institute for Theoretical Physics, California Institute of Technology, Pasadena CA, USA}
\affiliation{Joint Center for Quantum Information and Computer Science, National Institute of Standards and Technology \& University of Maryland, College Park MD, USA}

\date{\today}

\begin{abstract}

The Rabi model describes the simplest nontrivial interaction between a few-level system and a bosonic mode, featuring in multiple seemingly unrelated systems of importance to quantum science and technology. While exact expressions for the energies of this model and its few-mode extensions have been obtained, they involve roots of transcendental functions and are thus cumbersome and unintuitive. Utilizing the symmetric generalized rotating wave approximation (S-GRWA), we develop a family of approximations to the energies of the two-mode two-photon Rabi model. The simplest elements of the family are analytically tractable, providing better approximations in regimes of interest than the RWA such as in the ultra- and deep-strong coupling regimes of the system. Higher-order approximate energies can be used if more accuracy is required.

\begin{description}
\item[Keywords]
Rabi Hamiltonian, two bosonic modes, two-level systems, rotating wave approximation,\\
approximate spectrum
\end{description}
\end{abstract}

\maketitle

\section{\label{sec:level1}Introduction}



Introduced 80 years ago, the Rabi model has proven successful in describing the most basic interaction between light and matter [\onlinecite{Rabi}]. The original model depicts the interaction between an atom, approximated by a two-level system, and a single mode of radiation. Due to its simplicity and ubiquitousness, the model has found applications in various fields ranging from quantum optics  [\onlinecite{Deutsch2000,Vedral,rabl2009,macquarrie2013}] to solid state physics [\onlinecite{spin_boson_shore_sander_fg,WagnerBook,Irish,lewenstein2006,tomka2015,bernevig2006}]. In particular, its prominent presence in Josephson junctions [\onlinecite{SornborgerClelandGeller}], transmons \cite{Koch2007}, and flux qubits [\onlinecite{FornDiaz}] has cemented the model, and its simplified Jaynes-Cummings and dispersive versions \cite{jc}, as a cornerstone of quantum technology. 

Interest in the model and its extensions to multi-level systems and multiple modes continues, as such systems have presented many exciting experimental opportunities to explore rich physical phenomena \cite{DongZhangAmini,mr2019,clo2012,hpp2015,m2020,PhysRevA.95.053854,Boite,FornDiaz19,PhysRevA.99.022302,PhysRevA.99.063828,PhysRevA.92.033817,PhysRevA.98.053859,PhysRevA.97.013851}. In particular, cavity and circuit QED experiments have proposed constructions of quantum switches [\onlinecite{Zhou08,Zhou09,Thompson,Wallraff,Astafiev}], single-photon transistors [\onlinecite{Witthaut}], and various two-qubit gates [\onlinecite{Chow},\onlinecite{SchmidtKaler}]. Stabilization schemes for quantum computing require three- or higher-wave mixing terms \cite{Leghtas2014}. Moreover, tuning the light-matter coupling provides new avenues for exploring controlled photon entanglement experiments [\onlinecite{Maunz}] in two-photon quantum systems.

All of the mentioned efforts drive the necessity to better understand Rabi-type models analytically. Despite its simple looks, this has actually proven quite hard for even the original case [\onlinecite{Pellizzari}].
Early work [\onlinecite{EmaryBishop}] utilized the parity symmetry of the model to derive Juddian solutions --- analytic expressions for energies at crossings between eigenstates of different parity [\onlinecite{Judd77},\onlinecite{Judd79}]. Only recently, however, has Braak's use of discrete symmetries obtained the spectrum of the model, deeming it ``exactly solvable'' [\onlinecite{Braak}]. In the ``exact'' solution, eigenvalues of the Rabi Hamiltonian are zeroes of an infinite series of transcendental functions. While such solutions demonstrate a novel notion of integrability and are a mathematical step forward, their complexity presents a formidable obstacle to increasing physical understanding.

Recently, there has been additional interest in finding analytical solutions to generalizations of the quantum Rabi model \cite{Zhang13,AlbertScholesBrumer,pub009,cr15,xie20,zxgbgl2014,da2014,aa2019,rhd2019,PhysRevA.85.043805,DuanXieBraakChen,PhysRevA.99.013815,PhysRevA.103.023707}. These include the two-photon Rabi model, where the coupling between the spin and the oscillator creates photons in pairs, and the two-mode two-photon Rabi model, where each photon in the pair corresponds to a different oscillator. 
These models (listed in Table \ref{table:sym}) can be thought of as belonging to the same ``integrable'' \cite{Braak} class as the original Rabi model, and ``exact'' solutions of similar complexity have also been derived for them using different mathematical tools \cite{Zhang13,Zhang17,da2014,dhbc2015,cr15,xie20}. However, physically motivated yet accurate approximation methods, such as generalizations of the rotating wave approximation \cite{Irish,AlbertScholesBrumer}, have not yet been fully applied to these cases.

In this paper, we apply the symmetric generalized rotating wave approximation (S-GRWA) \cite{AlbertScholesBrumer} to the two-mode two-photon Rabi model, deriving a closed-form and relatively simple expression for the energies of the system. This approximation respects the underlying composite symmetry of the system and remains accurate in regimes of large spin-oscillator coupling. The approximation accuracy is tunable and is governed by how many matrix elements of a partially diagonalized Hamiltonian one includes in the approximation. We verify our approximation method against exact numerical results.

In Section \ref{sec:sgrwa}, we briefly review the S-GRWA. In Section \ref{sec:twomodetwophoton}, we apply the S-GRWA to the two-mode two-photon system.

\begin{table*}
\begin{tabular}{llll}
\hline\hline
\toprule 
Rabi-type model $H$ & $\Z_{2}$ symmetry $\sigma_x \cal{P}$ ~~~ & Additional symmetry~~~ & S-GRWA unitary\tabularnewline
\hline\midrule
\midrule 
$\omega\ph+J\sigma_{x}+\lambda\sigma_{z}(a+a^{\dagger})$ & $\sigma_{x}e^{i\pi\ph}$ & none & $\exp[\frac{\lambda}{\omega}(a-a^{\dagger})\sigma_{z}]$\tabularnewline
$\omega\ph+J\sigma_{x}+\lambda\sigma_{z}(a^{2}+a^{2\dagger})$ & $\sigma_{x}e^{i\pi{\ph \choose 2}}$ & $e^{i\pi\ph}$ ($\Z_{2}$) & $\exp[\alpha(a^{2}-a^{2\dagger})\sigma_{z}]$\tabularnewline
$\omega(\ph+\phm)+J\sigma_{x}+\lambda\sigma_{z}(a\bb+a^{\dagger}\bb^{\dagger})$~~~~ & $\sigma_{x}e^{i\pi{\ph+\phm \choose 2}}$ & $\ph-\phm$ ($\U(1)$) & $\exp[\alpha(a b-a^{\dagger}b^{\dagger})\sigma_{z}]$\tabularnewline
\bottomrule
\hline\hline
\end{tabular}

\caption{The three Rabi-type models covered by the S-GRWA approximation, their
symmetries, and the unitary operators used in the approximation procedure.
The first two models were treated in Ref.~[\onlinecite{AlbertScholesBrumer}], while the third one is treated here.}
\label{table:sym}
\end{table*}

\section{\label{sec:sgrwa}Generalized Symmetric Rotating Wave Approximation}

The rotating wave approximation (RWA) has been a well established method for solving the Rabi model, a well known Hamiltonian in quantum optics. This approximation works well when the assumption that the coupling strength between the two systems is weak such that the Stark effect and Bloch-Siegart shift are small [\onlinecite{ShoreKnight}]. The RWA works on the basis that fast oscillating terms can be ignored [\onlinecite{Schleich}]. However, in doing this, the RWA is not as effective when solving generalized Rabi-type models, such as the two-photon or two-mode models, as it fails to deliver accuracy in the strongly coupled regions of the Hamiltonian \cite{ToorZubairy,MilonniAckerhaltGalbraith,Lo2014,Casanova}. Therefore, a more powerful method is required for solving such systems.

Let us review the steps leading to the symmetric generalized rotating wave approximation (S-GRWA)  [\onlinecite{AlbertScholesBrumer}] in a way that applies to all models that we consider. We consider Rabi-type systems coupling a two-level system to one or two bosonic modes. Their Hamiltonians are all of the form 
\begin{equation}
H=\hb+J\sigma_{x}+\sigma_{z}V_{\text{bos}}\,,\label{eq:Ham}
\end{equation}
with $\hb$ and $\vb$ both purely bosonic. The former operator represents the purely bosonic ``frequency'' of the Hamiltonian, while $\vb$ is the bosonic part of the spin-boson coupling. The spin operators $\sigma_z,\sigma_x$ are the usual Pauli matrices, while $J$ is a parameter.

Writing out the Hamiltonian (\ref{eq:Ham}) explicitly in the $\sigma_z$ eigenstate basis of the spin yields a two-by-two matrix consisting of bosonic operators,
\begin{equation}
\label{eq:exactH}
H=\begin{pmatrix}\hb+\vb & J\\
J & \hb-\vb
\end{pmatrix}\,.
\end{equation}
The next step is to partially diagonalize this Hamiltonian in the spin space, which is done with the Fulton-Gouterman transformation \cite{fg_original}. This results in the removal of
off-diagonal entries in the above matrix. The transformation can be thought of as a
spin Hadamard rotation ``conditional'' on the state of the boson,
\begin{equation}
U=\frac{1}{\sqrt{2}}\begin{pmatrix}1 & -{\cal P}\\
{\cal P} & 1
\end{pmatrix}\,,\label{eq:FG}
\end{equation}
where the bosonic \textit{parity} operator ${\cal P}$ satisfies
\begin{equation}
\left[{\cal P},\hb\right]=0\,\,\,\,\,\,\,\,\,\,\,\,\,\text{and}\,\,\,\,\,\,\,\,\,\,\,\,\,{\cal P}^{2}=1\,.\label{eq:parityone}
\end{equation}
Applying this transformation to $H$ and once again writing out the
spin part, one will see that the spin off-diagonal entries
in $U^{\dg}HU$ are proportional to the anticommutator of $\vb$ and
${\cal P}$. Thus, in order to remove the off-diagonal entries,
the parity operator needs to anticommute with the coupling,
\begin{equation}
\vb{\cal P}=-{\cal P}\vb\,.\label{eq:paritytwo}
\end{equation}
If one can find a ${\cal P}$ satisfying properties (\ref{eq:parityone},\ref{eq:paritytwo}), then one obtains
the spin block-diagonal Hamiltonian
\begin{equation}
\label{eq:spinDiag}
U^{\dg}HU=\begin{pmatrix}\hb+\vb+J{\cal P} & 0\\
0 & \hb-\vb-J{\cal P}
\end{pmatrix}.
\end{equation}

For the single-mode Rabi model, $\vb\propto a+a^{\dg}$ (with $a$ being the mode's annihilation operator),  ${\cal P}=e^{i\pi a^{\dg}a}$
is simply the parity operator. For the two-mode case with coupling
$\vb\propto a^{2}+a^{\dg2}$, Ref.~\cite{AlbertScholesBrumer} introduced a generalized
parity (see Table~\ref{table:sym}) with which they were able to block-diagonalize
that case. Here, we introduce a two-mode parity operator with which
we block-diagonalize the two-mode two-photon Rabi model with coupling
$\vb\propto ab+a^{\dg}b^{\dg}$. This partial diagonalization was implicitly done in Ref. [\onlinecite{Zhang13}], but here we give the explicit form of the parity on the entire Hilbert space. 

Once the Hamiltonian is diagonal in the spin degree of freedom, we are left with two bosonic Hamiltonians $H^{\pm}=\hb+\vb\pm J{\cal P}$ containing the nonlinear component ${\cal P}$. Additional symmetries in the latter two-photon Rabi models can then be used to further block diagonalize the Hamiltonians in the bosonic space (see Table~\ref{table:sym}). This procedure exhausts all symmetries present in the system and brings about substantial numerical advantages: we are left with bosonic Hamiltonians that correspond to tridiagonal matrices in the bosonic Fock space.

Instead of attempting to diagonalize the resulting tridiagonal matrices exactly, the S-GRWA proceeds by applying another unitary rotation to each matrix, and then truncating the resulting matrix into block-diagonal form. For example, in the case of the Rabi model, a conditional (i.e., with direction depending on the spin state) displacement operator $\cal D$ is applied, and the resulting matrix of Laguerre polynomials (matrix elements of the displaced parity operator ${\cal D^\dg P D} = {\cal PD}^2$) is truncated into blocks of varying sizes. For the other two systems, squeezing operators are used instead of displacement. This forced block-diagonalization may seem ad-hoc, but it is rooted in effectively similar procedures in the original RWA and its generalization, the GRWA \cite{Irish}. It was verified that such truncations reproduce approximations obtained via other means \cite{AlbertScholesBrumer,AmniatTalab,liu}. The other benefit gained from taking this leap of faith is that, unlike the GRWA, the S-GRWA is better aligned with the $\Z_{2}$ symmetry $\sigma_x \cal{P}$ of the joint system than the other approximations \cite{AlbertScholesBrumer}.

\section{\label{sec:twomodetwophoton}Two-Mode Two-Photon System}
The Hamiltonian of the two-mode two-photon Rabi model, introduced in \cite{Zhang13,Zhang17,da2014,dhbc2015}, is
\begin{equation}
    H=\omega_1 a^{\dagger}a+\omega_2 b^{\dagger}b+J\sigma_x+\lambda\sigma_z(ab+a^{\dagger}b^{\dagger}),
\end{equation}
where $\lambda$ is the interaction strength, $\sigma_x,\sigma_z$ are the Pauli matrices describing the two-level system with energy gap $2J$, and $\omega_1,\omega_2$ are the frequencies of the bosonic modes with annihilation(creation) operators $a(a^{\dagger})$ and $b(b^{\dagger})$. The complete set of commutation relations among these operators is
\begin{align}
\begin{split}
    [a,b]=[a^{\dagger},b^{\dagger}]=0,&\quad [a,b^{\dagger}]=[a^{\dagger},b]=0,\\
    [a,a^{\dagger}]=1,&\quad [b,b^{\dagger}]=1.
\end{split}
\end{align}

Similar to the single-mode cases (see Table \ref{table:sym}), we first apply the Fulton-Gouterman transformation (\ref{eq:FG}) that block-diagonalizes the system in spin space using the parity operator
\begin{equation}
\label{eq:parityOperator}
    \mathcal{P}=\text{exp}\left[i\frac{\pi}{2}(a^{\dagger}a+b^{\dagger}b)(a^{\dagger}a+b^{\dagger}b-1)\right].
\end{equation}
Plugging this into the formulas of the previous section transforms the Hamiltonian into a diagonal matrix in spin space with bosonic entries
\begin{equation}
    H^{\pm}=\omega_{1}a^{\dagger}a+\omega_{2}b^{\dagger}b\pm\lambda(ab+a^{\dagger}b^{\dagger})\pm J{\cal P}.\label{eq:hamintermediate}
\end{equation}
In this way, the spin degree of freedom has been decoupled.

\begin{figure}
    \centering
    \includegraphics[width=.48\textwidth]{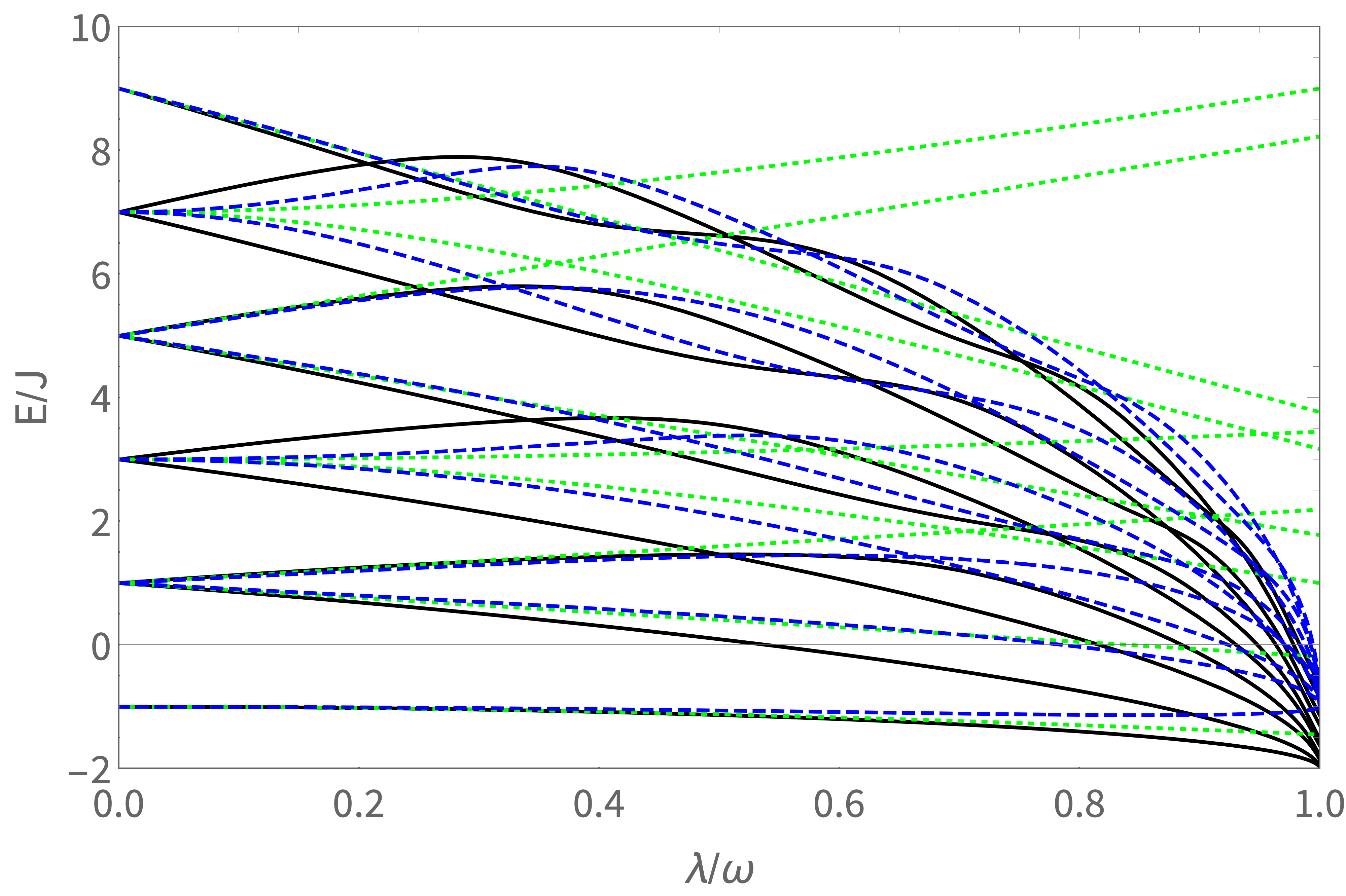}
    \caption{Comparison among eigenenergies of the two-mode two-photon Rabi Hamiltonian obtained by diagonalizing Eq. (\ref{eq:exactH}) truncated to a $100$-by-$100$ matrix (solid black lines), applying the RWA (dotted green lines), and applying the S-GRWA Eq. (\ref{eq:SGRWAH3}) and using its next-to-simplest eigenenergies obtained from diagonalizing Eq. (\ref{eq:SGRWAH322}) (dashed blue lines) for $\Delta=0,J=\omega=\omega_1=\omega_2=1$.}
    \label{fig:comparison}
\end{figure}

\begin{figure*}[htp]
\subfloat[$\Delta=0$]{%
  \includegraphics[width=0.33\textwidth]{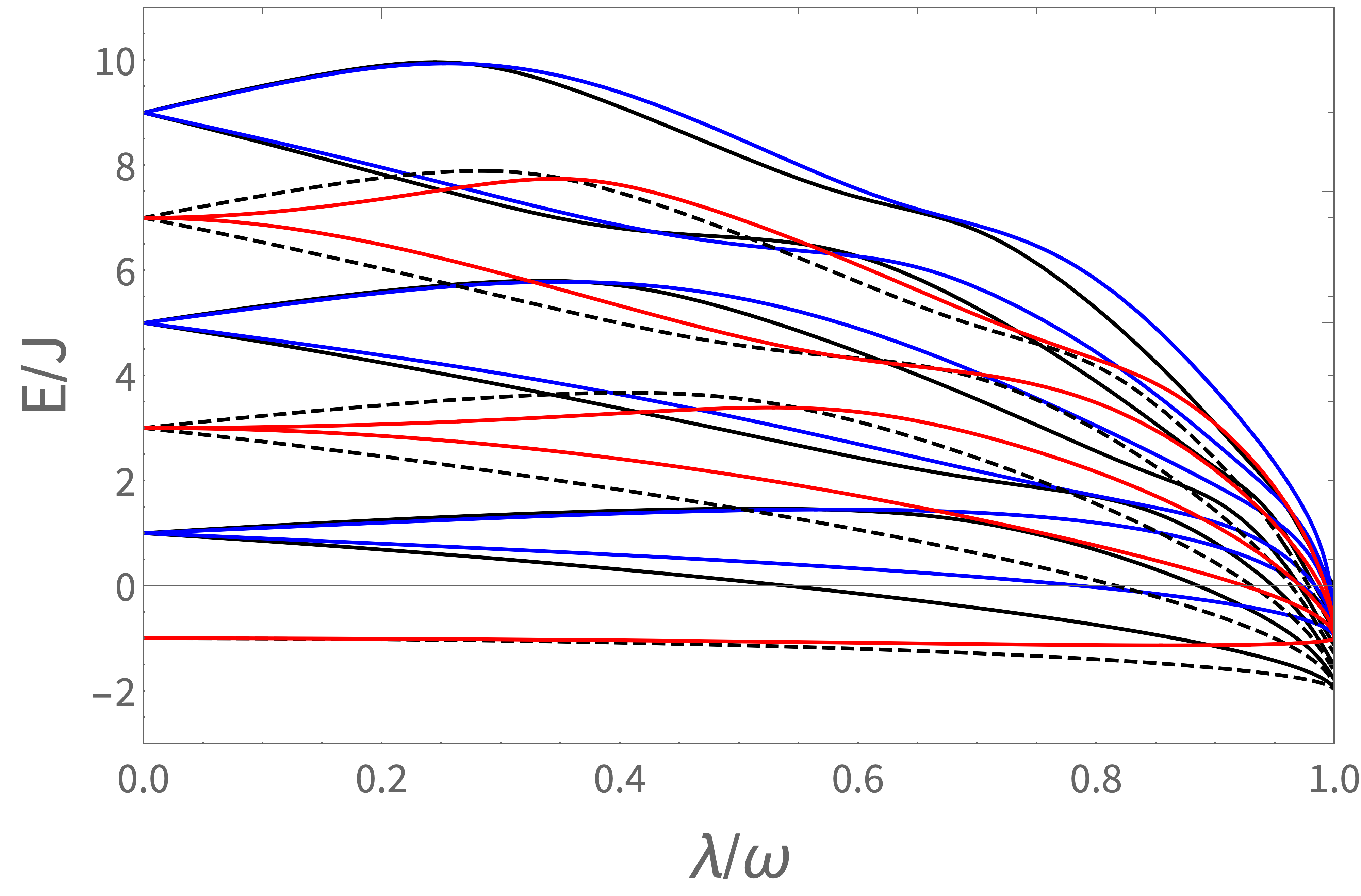}%
}
\subfloat[$\Delta=1$]{%
  \includegraphics[width=0.33\textwidth]{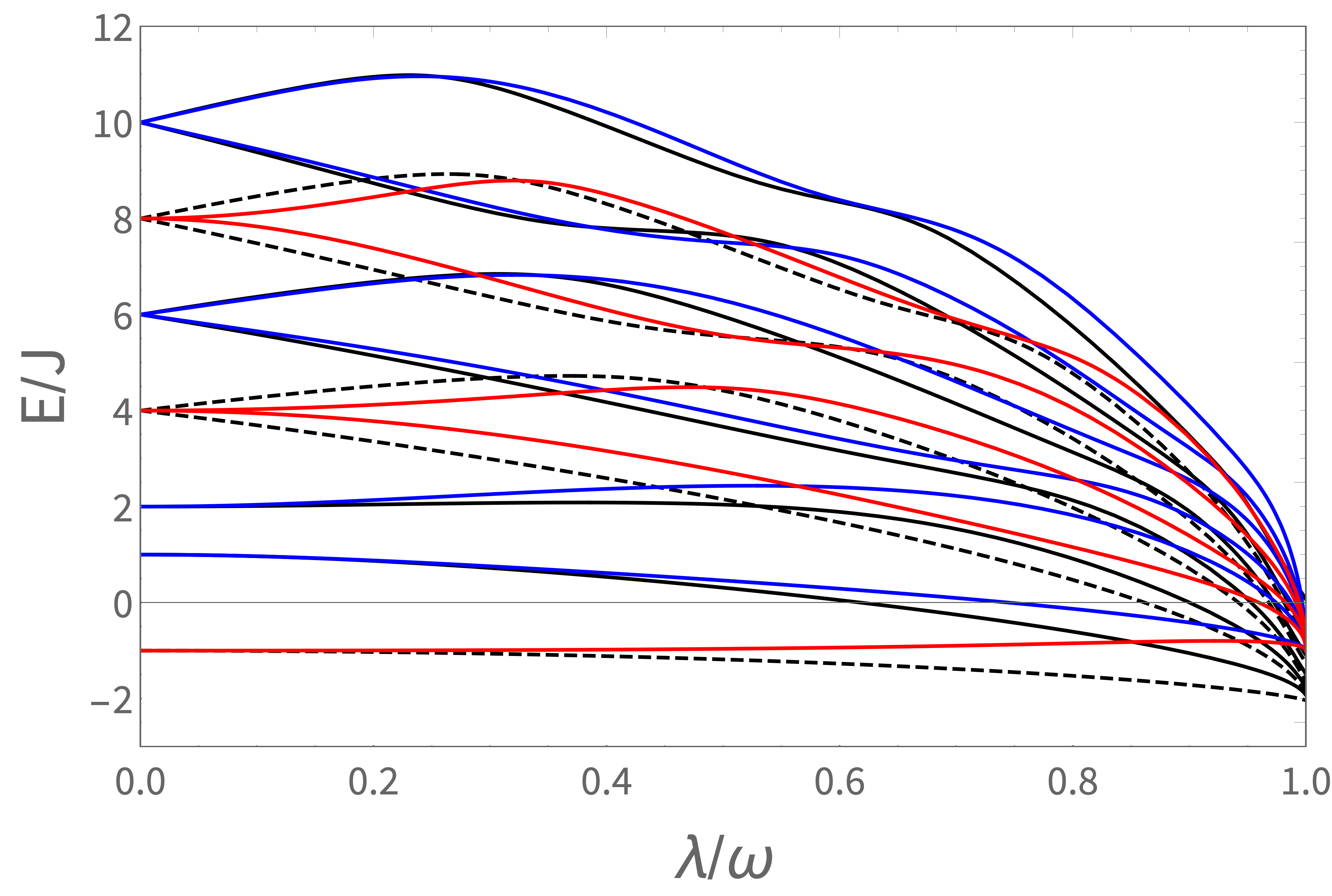}%
}
\subfloat[$\Delta=2$]{%
  \includegraphics[width=0.33\textwidth]{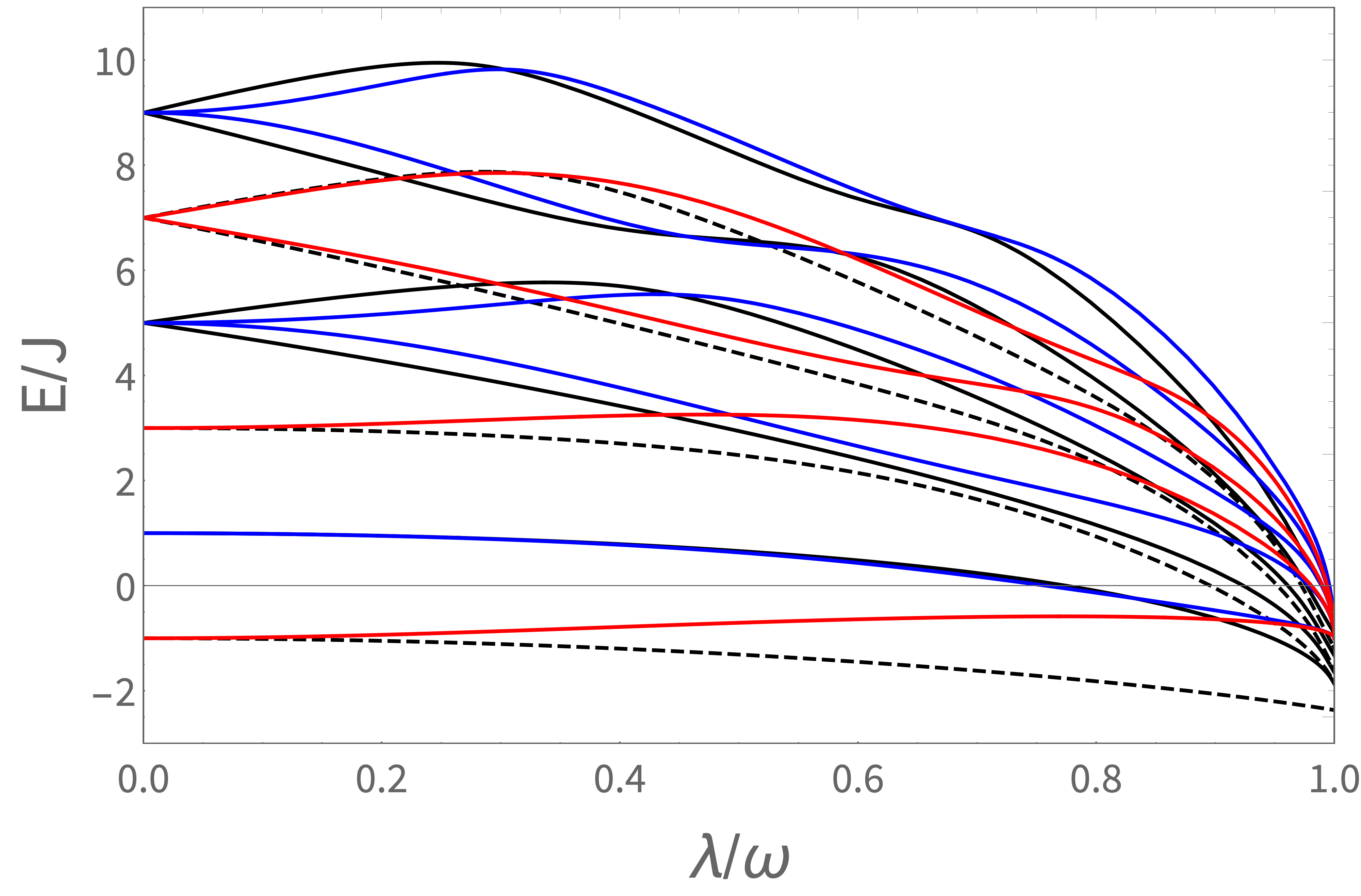}%
}

\subfloat[$\Delta=0$]{%
  \includegraphics[width=0.33\textwidth]{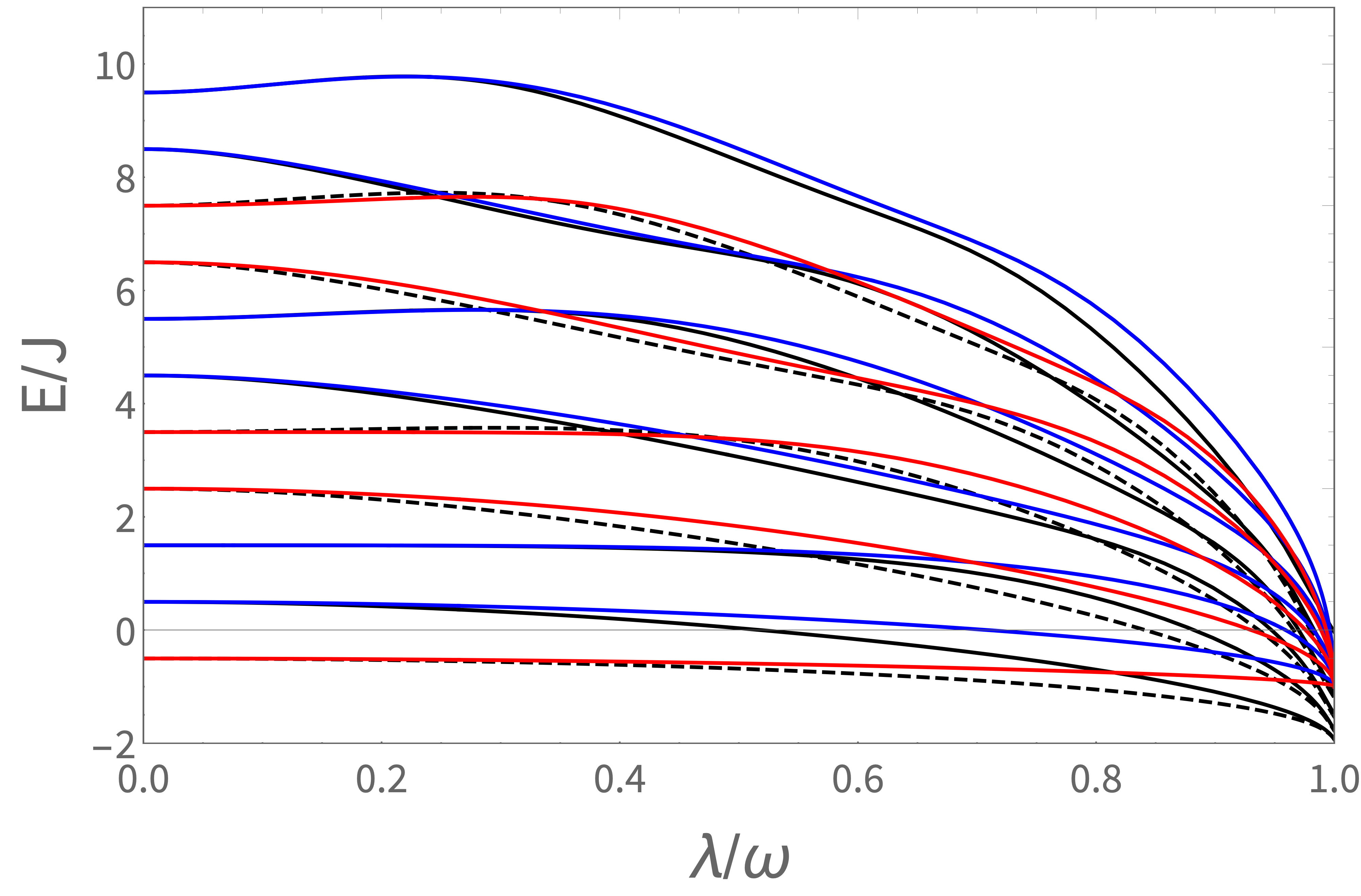}%
}
\subfloat[$\Delta=1$]{%
  \includegraphics[width=0.33\textwidth]{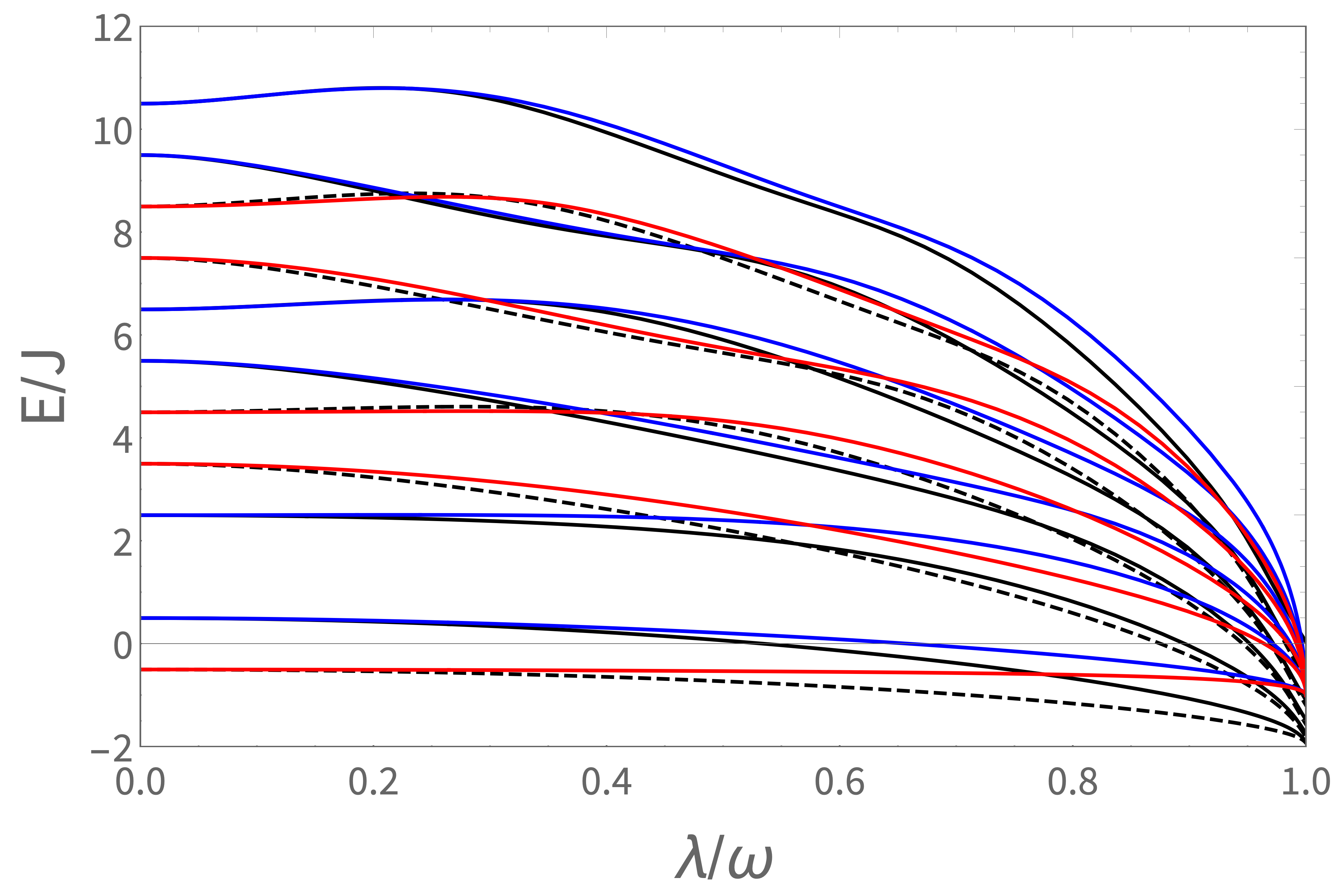}%
}
\subfloat[$\Delta=2$]{%
  \includegraphics[width=0.33\textwidth]{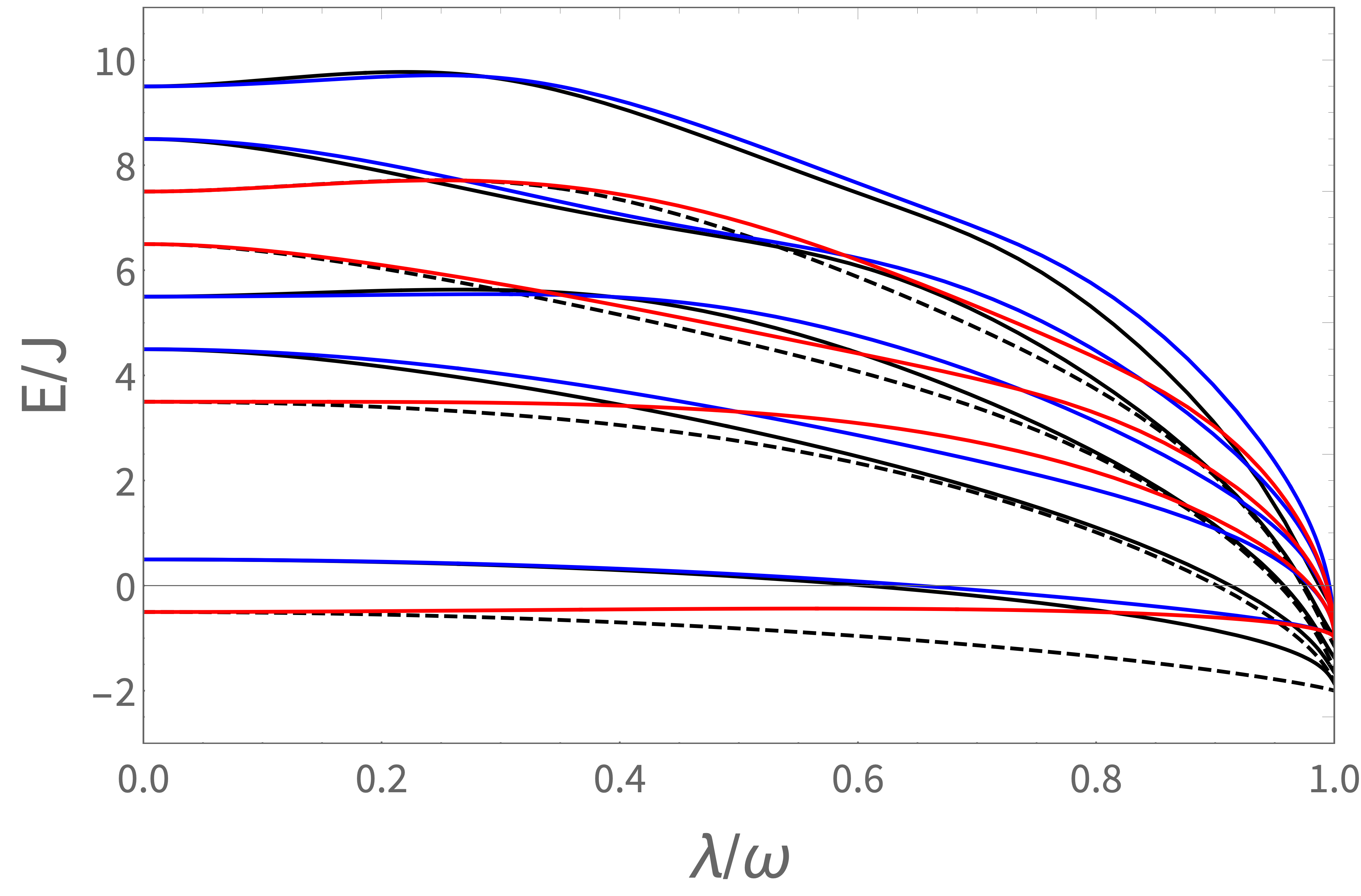}%
}
\caption{Exact solution of the two-mode two-photon Rabi Hamiltonian (black lines), plotted with the S-GRWA results (colored lines). The black lines represent the numerical solution corresponding to properly diagonalizing Eq. \eqref{eq:hamintermediate}'s tridiagonal matrices, each truncated to a 100-by-100 matrix in the Fock subspace of fixed photon number difference (\ref{eq:delta}). The positive (negative) parities of each manifold is represented via solid (dashed) black lines for exact energies and solid blue (red) lines for the S-GRWA energies. The results represent two cases: (a-c) $J=\omega=1$ and (d-f) $2J=\omega=1$ where $\omega=\omega_1=\omega_2$.
}
\label{fig:H3results}
\end{figure*}

Each of the above bosonic Hamiltonians has a $\boldsymbol{U}(1)$ symmetry, generated by the photon number difference operator 
\begin{equation}
    \hat\Delta = a^\dg a - b^\dg b.\label{eq:delta}
\end{equation}
As $H^{\pm}$ acts on the two-mode Fock space, where each state can be labeled as $\ket{n,m}$ for $n,m\geq0$, we can further decompose each bosonic Hamiltonian into block diagonal form,
\begin{equation}
\label{eq:subspaceDecomp}
    H^{\pm}=\bigoplus_{\Delta\in{\Z}}H_{\Delta}^{\pm}.
\end{equation}
Each Hamiltonian block in this finer form acts on its corresponding subspace of fixed photon number difference $\Delta = n-m$. The Hamiltonian on the block is a tridiagonal matrix when written in the fixed-$\Delta$ Fock subspace $\{|n,n+\Delta\rangle\}_{n=0}^\infty$ for $\Delta\geq 0$ (with the entries swapped for $\Delta < 0$ \cite{paircat}). The parity operator $\cal{P}$ earns its name by reducing to a diagonal operator proportional to $(-1)^n$ in this basis.

To complete the S-GRWA, we apply a similar two-mode squeezing operator used in [\onlinecite{da2014}] in order to get rid of the two-mode driving term $ab+a^\dg b^\dg$:
\begin{equation}
    \mathcal{S}(\pm\alpha)=\text{exp(}\pm\alpha[ab-a^{\dagger}b^{\dagger}]\text{)}.
\end{equation}
The sign of the squeezing will be conditional on the $\pm$ spin block, while the magnitude $\alpha>0$ satisfies
\begin{equation}
    \text{tanh}(2\alpha)=\lambda/\omega_{+},
\end{equation}
where it becomes convenient to express frequencies as $\omega_\pm = \frac{1}{2}(\omega_1 \pm \omega_2)$ from now on. Upon squeezing, the bosonic Hamiltonians transform as \cite{WagnerBook}
\begin{equation}
\label{eq:SGRWAH3}
    H_{\Delta}^{\pm}\to\tilde{\omega}(a^{\dagger}a+b^{\dagger}b)\pm J\mathcal{P}\mathcal{S}(\pm2\alpha),
\end{equation}
with a new $\lambda$-modulated frequency
\begin{equation}
    \tilde{\omega}=\sqrt{\omega_{+}^{2}-\lambda^{2}}
\end{equation}
and an unimportant constant offset $\tilde{\omega}-\omega_{+}+\omega_{-}\Delta$ that we have removed (although one should keep in mind that the $\Delta$ part of the offset is not constant on the entire Hilbert space). We can readily read off the regime of physicality of the system from the equation for $\tilde{\omega}$. When $\lambda$ surpasses $\omega_+$, the bosonic frequency becomes negative and the discrete spectrum collapses. This is readily seen in the numerical plots in Figure \ref{fig:H3results}. The collapse of the spectrum was also predicted in \cite{longliu2000,dhbc2015,da2014}.

So far, the transformations have yielded exact results. We now express the above Hamiltonians in the Fock subspace basis $\{|n,n+\Delta\rangle\}_{n=0}^\infty$ and further truncate to a direct sum of blocks of fixed size. The simplest truncation results in simply taking the diagonal entries (i.e., one-by-one blocks), which yields the following approximate energies,
\begin{equation}
    \tilde{\omega}(2n+\Delta)\pm J{(-1)}^{n}\sech^{\Delta+1}(\alpha) ~ P_{2n}^{(\Delta,0)}\left(2\tanh^{2}\alpha-1\right),
\end{equation}
indexed by $\pm$, Fock state index $n$, and photon number difference $\Delta$. Above, we have plugged in the diagonal $m=n$ matrix elements of the squeezing operator, 
\begin{equation}
    S^{\Delta}_{m,n}=\bra{m,m+\Delta}\mathcal{S}(2\alpha)\ket{n,n+\Delta},
\end{equation}
written in terms of Jacobi polynomials $P^{(\alpha,\beta)}_{n}(x)$.

For the more accurate two-by-two block truncation, we require off-diagonal matrix elements of the two-mode squeezing operator  [\onlinecite{Dariano}] (see also [\citealp{Ja}, Sec.~6.5.6]):
\begin{multline}
    S^{\Delta}_{m,n}
    =(-1)^{n}\eta^{m-n}(1-|\eta|^2)^{\kappa}\sqrt{\frac{n!\Gamma(2\kappa+m)}{m!\Gamma(2\kappa+n)}}\\
    \times \Psi_n^{2\kappa-1,m-n}(|\eta|^2)
\end{multline}
given that $\eta=\text{tanh}\alpha$, $2\kappa-1=\Delta$, and 
\begin{multline}
\label{eq:jacobi2p2m}
\Psi_n^{\beta,m-n}(\rho)\\=
\begin{cases}
P_n^{(\beta,m-n)}(2\rho-1) &\text{for }m\geq{n}\\
P_m^{(\beta,n-m)}(2\rho-1)\rho^{n-m}\frac{m!\Gamma(\beta+n+1)}{n!\Gamma(\beta+m+1)}  &\text{for }m\leq{n}.
\end{cases}
\end{multline}

We truncate the Hamiltonian into two-by-two blocks such that the $n$th block is
\begin{equation}
\label{eq:SGRWAH322}
\!\!
\begin{pmatrix}
\tilde{\omega}(2n+\Delta)\pm JS_{2n,2n}^{\Delta} & JS_{2n,2n+2}^{\Delta}\\
JS_{2n+2,2n}^{\Delta} & \!\!\!\!\tilde{\omega}(2n+2+\Delta)\mp JS_{2n+2,2n+2}^{\Delta}
\end{pmatrix}
\end{equation}
Diagonalizing this matrix yields the S-GRWA energies for the two-mode two-photon Rabi Hamiltonian. One can consider higher-dimensional blocks if more accuracy is required.

The accuracy of this two-by-two block approximation is depicted above in both \figref{fig:comparison} and \figref{fig:H3results}. In \figref{fig:comparison}, we present the eigenenergies obtained using exact diagonalization, the RWA, and the S-GRWA defined previously. The RWA eigenenergies are obtained from applying the unitary parity operator and, similar to our S-GRWA, we perform a two-by-two matrix truncation to derive the RWA eigenenergies numerically. We can see that for the first couple of eigenenergies of the system, both the RWA and the S-GRWA provide reasonable approximations in the weak-coupling regime $\lambda\ll \omega$. However, we note that, at resonance ($J=\omega_1=\omega_2$), relative to the exact eigenenergies of the system, the S-GRWA eigenenergies are more accurate in this regime than the standard RWA eigenenergies. Furthermore, the RWA fails to deliver accuracy in the ultra- and deep-strong coupling regime for higher eigenenergies of the system and exhibits significant divergence from the exact eigenenergies. In particular, while the S-GRWA manages to capture the spectral collapse of the system, the RWA fails to do so. Nevertheless, despite exhibiting the proper scaling and level crossings as they approach the point of collapse, the simplest S-GRWA energies and the point of spectral collapse are both slightly offset form the exact results. Additionally, we note that, particular to the ground state energy, the RWA remains more accurate than the S-GRWA even at higher-order approximations.

In \figref{fig:H3results}, the comparison is done by comparing the approximated energies obtained from the S-GRWA against the exact energies obtained by exact diagonalization of Eq. \eqref{eq:exactH}. In particular, we are concerned with two cases: (1) $J=\omega_1=\omega_2=1$, the resonance case, and (2) $2J=\omega_1=\omega_2=1$, an off-resonance case. As we have the Hamiltonian consisting of infinite orthogonal Fock subspaces, we investigate the performance of this S-GRWA when $\Delta=0,1,2$ for each parity ($\pm$), i.e. the accuracy of S-GRWA in each $H^{\pm}_{\Delta}$ as described in Eq. \eqref{eq:subspaceDecomp}. We have that $H^-_\Delta$ contains the ground state in each fixed-$\Delta$ sector. Furthermore, the orthogonality of subspaces in the Hilbert space supports the argument of given an initial state in the positive manifold, then the resulting ground energy will be present strictly in the positive manifold's spectrum. Contrary to the observation from single-mode Rabi Hamiltonians in [\onlinecite{AlbertScholesBrumer}], the S-GRWA approach provides great accuracy at small values of $\lambda$ in the ultra-strong coupling regime while it becomes less accurate at larger values of $\lambda$ for lower eigenenergies of the entire spectrum of the Hilbert space, particularly the deep-coupling regime $\lambda \to \omega$. However, consider larger block truncations extends the accuracy of the energies into that regime. The method also provides a good approximation to the level crossings between energies associated with different values of $\Delta$.

\section{\label{sec:level1}Conclusion}
Building on previous results, we have shown that the symmetric general rotating wave approximation (S-GRWA) [\onlinecite{AlbertScholesBrumer}] that has been utilized to approximate the spectrum of single-mode two-photon Rabi Hamiltonian can be successfully extended into the two-mode two-photon quantum system. Using a similar approach, we have derived a parity operator that explicitly diagonalizes the spin part of the Hilbert space. Furthermore, we utilize the $\boldsymbol{U}(1)$ symmetry of the two-mode case to produce a finer block-diagonalization that is amenable to a numerical treatment. We then approximate the exact energies of the system via the S-GRWA. The derived eigenenergies can accurately capture the dynamics of the system even in complicated setups where counter-rotating terms play crucial effects in the interactions.

Using this approach, we can express the approximate spectrum of the two-level two-photon system in relatively simple form.

\begin{acknowledgments}
D.H.W. is supported by the Caltech and Dr.~Jane Chen SURF fellowship, Caltech Student-Faculty Programs. We gratefully acknowledge support from the Walter Burke Institute for Theoretical Physics at Caltech. The Institute for Quantum Information and Matter is an NSF Physics Frontiers Center. Contributions to this work by NIST, an agency of the US government, are not subject to US copyright. Any mention of commercial products does not indicate endorsement by NIST.
\end{acknowledgments}

\bibliography{ref}

\end{document}